\documentclass[twocolumn,showpacs,preprintnumbers,amsmath,amssymb]{revtex4-1}

\usepackage{graphicx}
\usepackage{dcolumn}
\usepackage{bm}
\usepackage{hyperref}

\begin{document}


\title{Transitions among the diverse oscillation quenching states induced by the interplay of direct and indirect coupling} 


\author{Debarati Ghosh}
\affiliation{Department of Physics, University of Burdwan, Burdwan 713 104, West Bengal, India.}
\author{Tanmoy Banerjee}
\thanks{The author to whom correspondence should be addressed: tbanerjee@phys.buruniv.ac.in}
\affiliation{Department of Physics, University of Burdwan, Burdwan 713 104, West Bengal, India.}

\received{:to be included by reviewer}
\date{\today}

\begin{abstract}
We report the transitions among different oscillation quenching states induced by the interplay of diffusive (direct) coupling and environmental (indirect) coupling in coupled identical oscillators. This coupling scheme was introduced by Resmi {\it et al}. [Phys. Rev. E, 84, 046212 (2011)] as a general scheme to induce amplitude death (AD) in nonlinear oscillators. Using a detailed bifurcation analysis we show that in addition to AD, which actually occurs only in a small region of parameter space, this coupling scheme can induce other oscillation quenching states, namely oscillation death (OD) and a novel nontrvial AD (NAD) state, which is a nonzero {\it bistable} homogeneous steady state; more importantly, this coupling scheme mediates a transition from AD to OD state and a new transition from AD to NAD state. We identify diverse routes to the NAD state and map all the transition scenarios in the parameter space for periodic oscillators. Finally, we present the first experimental evidence of oscillation quenching states and their transitions induced by the interplay of direct and indirect coupling. 
\end{abstract}

\pacs{05.45.Xt}
\keywords{Environmental coupling, Amplitude death, Oscillation death, Bistability, Nonlinear electronic circuit}

\maketitle 

\section{Introduction}
\label{sec:intro}
The suppression of oscillation is an important topic of research in the context of coupled oscillators and has been studied in diverse fields such as physics, biology, and engineering \cite{kosprep}. Two distinct types of oscillation quenching processes are there, namely  amplitude death (AD) and oscillation death (OD). The AD state is defined as a stable homogeneous steady state (HSS) that arises in coupled oscillators under some parametric conditions \cite{prasad1},\cite{asen,*asen1,*prasad3}; in the case of AD, oscillation is suppressed as all the coupled oscillators attain a common steady state that was unstable in the uncoupled condition. In the case of OD, oscillators populate coupling dependent stable inhomogeneous steady states (IHSS) that are resulted from the symmetry breaking bifurcation; e.g., in the case of two coupled oscillators, in the OD state oscillation is suppressed and they attain two different newly created coupling dependent steady states. In the phase space OD may coexist with limit cycle oscillation.  While AD is observed and characterized as a control mechanism to suppress oscillation in Laser application \cite{laser}, neuronal systems \cite{bard}, electronic circuits \cite{tanchaosad}, etc., the OD on the other hand is relevant in many biological and physical systems such as synthetic genetic oscillator \cite{kosepl,*koschaos}, neural network \cite{curtu}, laser systems \cite{odphys}, etc  (see \cite{kosprep} for an elaborate review on OD).

In the earlier works on oscillation suppression, no clear distinction between AD and OD was emphasized until the pioneering research by  \citet{kosprl}, where it was shown that the AD and OD are two dynamically different phenomena, both from their origin and manifestation. \citet{qstr2}  observed OD in genetic network under phase-repulsive coupling with realistic biological parameters and proved its importance in biological network. Later, it has been reported that most of the coupling schemes, which were believed to induce AD only can induce OD, also: Ref.\cite{kospre} proved that the dynamic and conjugate coupling can induce OD; Refs. \cite{tanpre1,*tanpre2} and \cite{dana,*dana1} reported the occurrence of OD induced by mean-field and repulsive coupling, respectively. More significantly, Refs. \cite{kosprl,kospre,tanpre1,tanpre2,dana,dana1} show that these coupling schemes can induce an important transition phenomenon, namely the transition from AD to OD that resembles the Turing-type bifurcation in spatially extended systems, which is believed to have connection with the phenomenon of cellular differentiation \cite{cell}. Further, in the recent studies new oscillation quenching state \cite{tanpre1,*tanpre2} and also new routes to oscillation quenching states \cite{dana,*dana1} are reported. More recently, in Ref.~\cite{kurthpre} the important rigorous conditions for the onset of AD and OD in a system of identical Stuart-Landau oscillators has been reported. Thus, search for the transitions between different oscillation quenching states and identification of new oscillation quenching phenomena are an active area of recent research on coupled oscillators.

In this paper we report the occurrence of the AD to OD transition and a new transition from AD to a novel nontrivial amplitude death (NAD) state induced by the simultaneous presence of diffusive coupling (i.e. direct coupling) and environmental coupling (i.e. indirect coupling). To the best of our knowledge, the AD to NAD transition has not been observed earlier for any other coupling configuration. This direct-indirect coupling scheme was originally proposed by \citet{resmiad} as a general scheme for inducing AD in coupled oscillators (later extended for a network of oscillators \cite{resmi12}), and attracts immediate attention due to its ease of implementation and generality to induce AD in any synchronizable units. Although diffusive coupling is widely studied in the context of synchronization but environmental coupling is a less explored topic; it is particularly important in biological systems, e.g., populations of cells in which oscillatory reactions are taking place interact with each other via chemicals that diffuse in the surrounding medium \cite{kat}. Since the authors of Ref.\cite{resmiad} rely mainly on time integration of the dynamical equations and linear stability analysis, thus the complete dynamical features induced by this coupling scheme were not explored and only the AD state was identified and characterized there. 

In the present paper we employ a detailed bifurcation analysis to show that apart from AD, which actually occurs in a small zone of parameter space, there exists diverse oscillation suppression states, namely OD and a newly observed nontrivial AD (NAD) state. This NAD state has not only a nonzero homogeneous steady state but, more significantly, in this state the system becomes {\it bistable} (will be elaborated later in this paper). We explore the properties of the NAD state and identify three distinct routes to NAD. More importantly, we recognize different transition scenarios, e.g. AD to OD transition, and a novel transition from AD to NAD state. The importance of this AD-NAD transition lies in the fact that it gives the evidence of {\it direct transition} from the mono-stability (AD) to bistability (NAD) in dynamical system that may improve our understanding of the origin of bistability arises in biological processes \cite{brain,*brain2}\cite{ecoli}.  In this study we choose the paradigmatic Stuart-Landau and Van der Pol oscillators and derive the range of coupling parameters where the transitions are occurred. Finally, we report the first experimental evidence of the NAD state and the AD-NAD transition; also, we experimentally observe AD, OD and AD-OD transitions that support our theoretical findings.

\section{Generic oscillators with direct-indirect coupling}
\label{sec2}
\subsection{Direct-indirect coupled Stuart-Landau oscillators}
At first, we consider two identical Stuart-Landau oscillators interacting directly through diffusive coupling and indirectly through a common environment $s$, which is modeled as a damped dynamical system \cite{resmiad}\cite{resmienv,*tanenv}. Mathematical model of the coupled system is given by
\begin{subequations}
\label{system}
\begin{align}
\label{x}
\dot{x}_{1,2} &= P_{1,2}x_{1,2}-{\omega}y_{1,2}+d(x_{2,1}-x_{1,2})+\epsilon s,\\
\label{y}
\dot{y}_{1,2} &= {\omega}x_{1,2}+P_{1,2}y_{1,2},\\
\label{s}
\dot{s} &= -ks-\frac{\epsilon (x_1+x_2)}{2}.
\end{align}
\end{subequations}
Here, $P_i = 1-{x_i}^2-{y_i}^2$ $(i = 1,2)$. The individual Stuart-Landau oscillators are of unit amplitude and having eigenfrequency $\omega$. The diffusive coupling strength is given by $d$, and $\epsilon$ is the environmental coupling strength that controls the mutual interaction between the systems and environment. $k$ represents the damping factor of the environment ($k>0$). Eq.(\ref{system}) has a trivial homogeneous steady state (HSS), which is the origin $(0,0,0,0,0)$, and additionally two coupling dependent nontrivial fixed points $\mathcal{F}_{IHSS}\equiv(x^\dagger, y^\dagger, -x^\dagger, -y^\dagger, 0)$, where 
$x^\dagger = -\frac{\omega y^\dagger}{2d{y^\dagger}^2+{\omega}^2}$, $y^\dagger =\pm  \sqrt{\frac{(d-{\omega}^2)+\sqrt{d^2-{\omega}^2}}{2d}}$, and  $\mathcal{F}_{NHSS}\equiv(x^\ast, y^\ast, x^\ast, y^\ast, s^\ast)$, where 
$x^\ast = -\frac{k\omega y^\ast}{k{\omega}^2+{\epsilon}^2{{y^\ast}^2}}$, $y^\ast =\pm  \sqrt{\frac{({\epsilon}^2-2k{\omega}^2)+\sqrt{{\epsilon}^4-4k^2{\omega}^2}}{2{\epsilon}^2}}$ and $s^\ast = -\frac{\epsilon x^\ast}{k}$. The fixed points $\mathcal{F}_{IHSS}$ give the inhomogeneous steady states (IHSS); note that it depends only upon $d$ and independent of $\epsilon$ and  $k$;  {\it stabilization} of IHSS  results in OD. The fixed points $\mathcal{F}_{NHSS}$ represent {\it nontrivial homogeneous steady states} (NHSS), stabilization of which gives rise to a novel nontrivial amplitude death (NAD) state, which is a nonzero bistable state. It can be seen that this NHSS depends upon $\epsilon$ and $k$, and independent of $d$.

The characteristic equation of the system at the trivial HSS, $(0,0,0,0,0)$, is given by,
\begin{equation}
\label{trieigen}
({\lambda}^2+P'_{T1}\lambda + P'_{T0})({\lambda}^3+P_{T2}{\lambda}^2+P_{T1}\lambda +P_{T0})=0,
\end{equation}
where $P'_{T1}=2(d-1)$, $P'_{T0}=(1-2d+{\omega}^2)$, $P_{T2}=(k-2)$, $P_{T1}=(1+{\omega}^2+{\epsilon}^2-2k)$ and $P_{T0}=(k+k{\omega}^2-{\epsilon}^2)$.

\begin{figure}
\includegraphics[width=.49\textwidth]{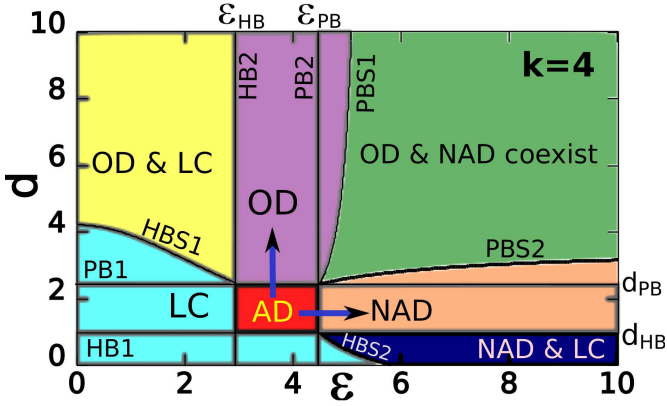}
\caption{\label{fig1} (Color online) Bifurcation diagram in $\epsilon-d$ space for Stuart-Landau oscillator using $\mbox{XPPAUT}$ ($\omega=2$ and $k=4$). Horizontal arrow shows AD-NAD transition and vertical arrow indicates AD-OD transition. AD occurs in a small region where ${\epsilon}_{HB}<\epsilon<{\epsilon}_{PB}$ and $d_{HB}<d<d_{PB}$ (for a detailed description see text).}
\end{figure}

Since Eq.\eqref{trieigen} is a fifth-order polynomial, it is difficult to predict bifurcation points from the eigenvalues; thus, we derive the bifurcation points from the properties of coefficients of the characteristic equation itself using the method stated in Ref.\cite{math}.  A close inspection of the nontrivial fixed points reveals that the IHSS, $\mathcal{F}_{IHSS}$, appears through a pitchfork bifurcation at 
\begin{equation}
d_{PB}=\frac{1+{\omega}^2}{2}.
\end{equation}
Also, the NHSS, $\mathcal{F}_{NHSS}$, appears through a pitchfork bifurcation at 
\begin{equation}\label{pbeps}
\epsilon_{PB} = \sqrt{k(1+{\omega}^2)}. 
\end{equation}
Further, from  Eq.\eqref{trieigen} we derive the following values of $d$ (by setting $P'_{T1}=0$ \cite{math}) and $\epsilon$ (by setting $P_{T1}P_{T2}-P_{T0}=0$ \cite{math}) where Hopf bifurcations (of trivial HSS) occur, respectively:
\begin{eqnarray}
d_{HB}&=&1,\label{hopfd}\\
\epsilon_{HB} &=& \sqrt{\frac{2(k-1)^2+2{\omega}^2}{k-1}}\label{hopfeps},
\end{eqnarray}
with an additional condition: $\omega>1$. We find that all the eigenvalues at the trivial HSS have negative real part and thus gives rise to AD when $d_{HB}<d<d_{PB}$, $k>2$, and ${\epsilon}_{HB}<\epsilon<{\epsilon}_{PB}$. To corroborate our analysis we compute the two parameter bifurcation diagram in $\epsilon-d$ space using the $\mbox{XPPAUT}$ package \cite{xpp}. Figure.\ref{fig1} shows this for $k=4$ (i.e. $k>2$)  and $\omega=2$: we observe that the parameter space is divided into different zones separated  by two horizontal lines at $d_{HB}$ and $d_{PB}$, and two vertical lines at $\epsilon_{HB}$ and $\epsilon_{PB}$. Also, in the $\epsilon-d$ parameter space the area of the rectangular zone where AD occurs is: $(d_{PB}-d_{HB})({\epsilon}_{PB}-{\epsilon}_{HB})$.

\begin{figure}
\includegraphics[width=.45\textwidth]{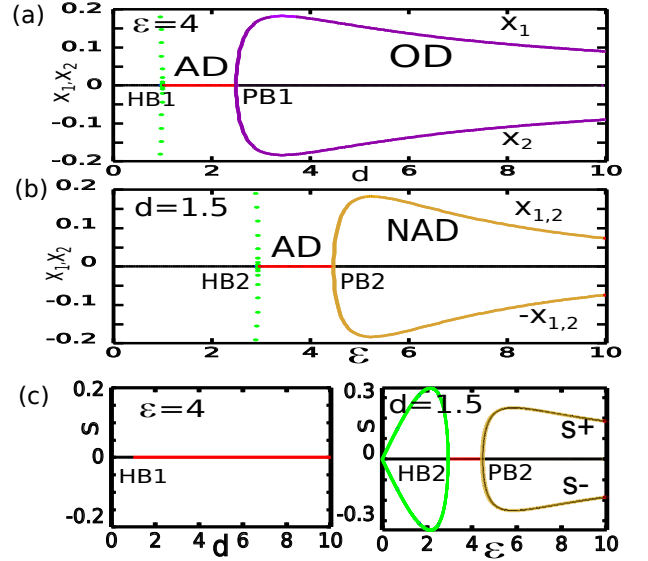}
\caption{\label{fig2} (Color online) (a) Transition from LC to AD  (at HB1), and AD to OD (at PB1) with the variation of $d$ ($\epsilon=4$, $k=4$). Black line: unstable steady state, deep gray (violet) line: OD state, light gray (golden) line: NAD state, (green) solid circles: stable limit cycle. (b) Transition from LC to AD  (at HB2), and AD to NAD (at PB2) with the variation of $\epsilon$ ($d=1.5$, $k=4$). Here, the NAD state is a bistable state: depending upon initial conditions one gets $x_{1,2}=\pm x^\ast$. (c) Variation of the environment, $s$: (left panel) in the OD state ($\epsilon=4$) $s$ attains the stable zero steady state beyond HB1; (right panel) between HB2 and PB2, $s$ attains the stable zero steady state but beyond PB2 the environment $s$ becomes bistable: depending upon initial conditions $s$ may attain either $s+$ or $s-$ state.}
\end{figure}

At this point we identify the following two distinct transition scenarios (for $k>2$): (i) If ${\epsilon}_{HB}<\epsilon<{\epsilon}_{PB}$, variation of $d$ gives rise to a transition from limit cycle (LC) to AD through Hopf bifurcation (HB1) at $d_{HB}$, and from AD to OD through pitchfork bifurcation (PB1) at $d_{PB}$ (see Fig.\ref{fig1}). This is also shown in one dimensional bifurcation diagram [Fig.\ref{fig2}(a)] at an exemplary value $\epsilon=4$ and $k=4$ where we see that AD occurs at $d_{HB}=1$ and OD emerges at $d_{PB}=2.5$; the OD state is represented by $x_1=-x_2=x^\dagger$.  (ii) If $d_{HB}<d<d_{PB}$, variation of $\epsilon$ induces a transition from LC to AD through Hopf bifurcation (HB2) at $\epsilon_{HB}$, and from AD to NAD through pitchfork bifurcation (PB2) at $\epsilon_{PB}$ (see Fig.\ref{fig1}); Fig.\ref{fig2}(b) shows this for $d=1.5$ and $k=4$: here AD occurs at $\epsilon_{HB}=2.943$ and NAD at $\epsilon_{PB}=4.472$. The direct transition from AD to NAD state is reported for the first time and has not been observed earlier for any other coupling scheme. The manifestation of NAD is interesting: now the system becomes {\it bistable}, i.e.,  depending upon the initial conditions it may attain two different steady states, either $x_1=x_2=x^\ast$ or $x_1=x_2=-x^\ast$. We also examine the change in environment in the above two cases. It is interesting to note that in the OD state since $x_1=-x_2$, the effect of the system on the environment vanishes [see Eq.\eqref{s}] and $s$ remains in its trivial zero steady state [Fig.\ref{fig2}(c), left panel]. But in the NAD state, since $x_{1,2}=\pm x^\ast$ the effect of the system on the environment is very much present there, thus like the NAD state the environment $s$ also becomes bistable: depending upon initial conditions it attains either $s+$ or $s-$ state beyond PB2 [Fig.\ref{fig2}(c), right panel].    
\begin{figure}
\includegraphics[width=.5\textwidth]{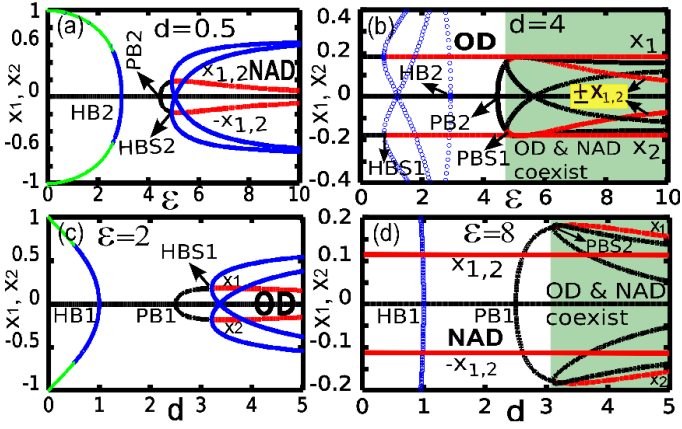}
\caption{\label{fig3} (Color online) (a) Variation of $\epsilon$ ($d=0.5$): The NAD state ($x_{1,2}=\pm x^\ast$) appears through subcritical Hopf bifurcation at HBS2. grey (red) line: stable steady state. Open (blue) circles: unstable limit cycle. (b) Variation of $\epsilon$ ($d=4$): The OD arises at HBS1 through subcritical Hopf bifurcation; a new NAD state emerges at PBS1 through subcritical pitchfork bifurcation and coexists with OD [gray (green) filled region]. (c) Variation of $d$ ($\epsilon=2$): The OD state appears through subcritical Hopf bifurcation at HBS1. (d) Variation of $d$ ($\epsilon=8$): The OD state appears through subcritical pitchfork bifurcation at PBS2 that coexists with the NAD state [gray (green) filled region].}
\end{figure}

Outside these horizontal and vertical rectangular regions [of width $(d_{PB}-d_{HB})$ and $({\epsilon}_{PB}-{\epsilon}_{HB})$, respectively] we identify several other bifurcation curves that mark the distinct regions of occurrence of oscillation suppression states and their coexistence. For this purpose at first we consider the characteristic equation corresponding to the fixed point $\mathcal{F}_{IHSS}\equiv(x^\dagger, y^\dagger, -x^\dagger, -y^\dagger, 0)$ which is given by
\begin{equation}
\label{ntrieigen}
({\lambda}^2+P'^{\dagger}_{N1}\lambda + P'^{\dagger}_{N0})({\lambda}^3+P^{\dagger}_{N2}{\lambda}^2+P^{\dagger}_{N1}\lambda +P^{\dagger}_{N0})=0,
\end{equation}
where $P'^{\dagger}_{N1}=2(d-1)+4{x^\dagger}^2+4{y^\dagger}^2$, $P'^{\dagger}_{N0}=(1-{x^\dagger}^2-3{y^\dagger}^2)(1-3{x^\dagger}^2-{y^\dagger}^2-2d)+{\omega}^2-4{x^\dagger}^2{y^\dagger}^2$, $P^{\dagger}_{N2}=4({x^\dagger}^2+{y^\dagger}^2)-2+k$, $P^{\dagger}_{N1}=1+{\omega}^2+{\epsilon}^2-2k+({x^\dagger}^2+{y^\dagger}^2)\{4k-4+3({x^\dagger}^2+{y^\dagger}^2)\}$. $P^{\dagger}_{N0}=k\{1-4({x^\dagger}^2+{y^\dagger}^2)+3({x^\dagger}^2+{y^\dagger}^2)^2+{\omega}^2\}-{\epsilon}^2(1-{x^\dagger}^2-3{y^\dagger}^2)$.
From Eq.\eqref{ntrieigen} we find that $\epsilon$ and $k$ appear only in the term $({\lambda}^3+P^{\dagger}_{N2}{\lambda}^2+P^{\dagger}_{N1}\lambda +P^{\dagger}_{N0})$, i.e., they control only three eigenvalues. 

We find the characteristic equation corresponding to the fixed point $\mathcal{F}_{NHSS}\equiv(x^\ast, y^\ast, x^\ast, y^\ast, s^\ast)$ as:
\begin{equation}
\label{nadeigen}
({\lambda}^2+P'^{\ast}_{N1}\lambda + P'^{\ast}_{N0})({\lambda}^3+P^{\ast}_{N2}{\lambda}^2+P^{\ast}_{N1}\lambda +P^{\ast}_{N0})=0,
\end{equation}
where $P'^{\ast}_{N1}$, $P'^{\ast}_{N0}$, $P^{\ast}_{N2}$, $P^{\ast}_{N1}$ and $P^{\ast}_{N0}$ are identical to the $P'$ and $P$ values of the above paragraph with the $\dagger$ sign is now replaced by the $\ast$ sign.

For $d<d_{HB}$, the NAD state appears through Hopf bifurcation at HBS2 (Fig.\ref{fig1}) which is obtained from Eq.\eqref{nadeigen} by putting $P'^{\ast}_{N1}=0$ \cite{math} and is given by
\begin{equation}
\label{slohbs2}
{\epsilon}_{HBS2}= \sqrt{\frac{k(d+1)^2+4k{\omega}^2}{2(d+1)}}.
\end{equation}
This is shown in Fig.\ref{fig3}(a) for $d=0.5$. We can see that the NAD state here coexists with an unstable limit cycle. 

For $d>d_{PB}$, three more bifurcation curves appear, namely HBS1, PBS1 and PBS2 (Fig.\ref{fig1}). We derive the locus of all the curves in the $\epsilon-d$ space. First, for $\epsilon<{\epsilon}_{HB}$, the OD appears through subcritical Hopf bifurcation at HBS1 whose locus is derived from Eq.\eqref{ntrieigen} by putting $P^{\dagger}_{N2}P^{\dagger}_{N1}-P^{\dagger}_{N0} = 0$ as,
\begin{equation}
\label{hbs1}
\epsilon_{HBS1} = \sqrt{\frac{2G_{12}[{\omega}^2+(k-1)^2+G_1(3G_1+4(k-1))]}{k-1+2{x^\dagger}^2+G_1}},
\end{equation}
with $G_1 = ({x^\dagger}^2+{y^\dagger}^2)$ and $G_{12}=(1-2G_1)$. This is shown  in Fig.\ref{fig3}(b) for $d=4$ ($d>d_{PB}$): between HBS1 and HB2 the OD state coexists with an unstable limit cycle, between HB2 and PBS1, OD is the only stable state. Beyond PBS1, OD  is accompanied by a NAD state that is created by a subcritical pitchfork bifurcation (at PBS1). 

For $\epsilon>{\epsilon}_{PB}$ (and $d>d_{PB}$), we have two subcritical pitchfork bifurcation curves, PBS1 and PBS2, locus of which are derived by putting $P'^{\ast}_{N0}=0$ and $P^{\dagger}_{N0}=0$ in Eq.\eqref{nadeigen} and Eq.\eqref{ntrieigen}, respectively:
\begin{subequations}
\label{slopbs1}
\begin{align}
d_{PBS1} &= \frac{1+G_2(3G_2-4)+{\omega}^2}{2(1-G_2-2{ y^\ast}^2)},\\
\epsilon_{PBS2} &= \sqrt{\frac{K[1+{\omega}^2+G_1(3G_1-4)]}{1-G_1-2{y^\dagger}^2}}.
\end{align}
\end{subequations} 
Here $G_2 = { x^\ast}^2+{ y^\ast}^2$. Between the curves PBS1 and PBS2, OD and NAD states coexist (see Fig.~\ref{fig1}). Note that bifurcation curves given by Eq.\eqref{slohbs2}--Eq.\eqref{slopbs1} depend upon both $d$ and $\epsilon$ (for a given $\omega$ and $k$). Figure \ref{fig3}(c-d) show the variation of the system dynamics for variable $d$. For $\epsilon=2$, OD emerges along with an unstable limit cycle at HBS1; for $\epsilon=8$ ($>\epsilon_{PB}$), NAD and LC coexists upto $d=d_{HB}$ (HB1), and between HB1 and PBS2, NAD is the only stable state, then beyond PBS2, NAD and OD coexist. Here the OD state emerges through subcritical pitchfork bifurcation (at PBS2). 

It is important to note that, unlike conventional AD that has only two routes, namely Hopf and saddle-node bifurcation \cite{prasad1}, we identify three distinct routes to NAD state: (i) {\it supercritical pitchfork bifurcation} route which occurs at $\epsilon_{PB}$ for $d_{HB}<d<d_{PB}$. (ii) {\it Subcritical pitchfork bifurcation} route (PBS1) that occur for $\epsilon>\epsilon_{PB}$ and $d>d_{PB}$. (iii) {\it Subcritical Hopf bifurcation} route that occurs for $d<d_{HB}$ along the HBS2 curve whose locus is given by Eq.\eqref{slohbs2}. Although the subcritical pitchfork bifurcation route to a different nontrivial AD state was observed earlier in the mean-field coupled oscillators \cite{tanpre1,*tanpre2}  the other two routes to NAD were not observed earlier for any other coupling schemes.

\begin{figure}
\includegraphics[width=.3\textwidth]{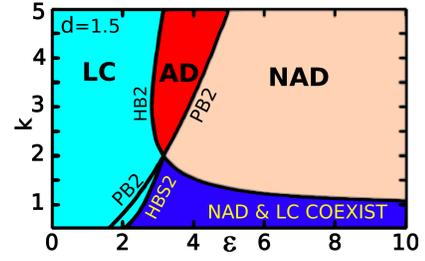}
\caption{\label{fig4} (Color online) Phase diagram in $\epsilon-k$ space ($d=1.5$). No AD (and thus AD-NAD transition) is possible for $k\le 2$.}
\end{figure}
We also examine the effect of environment (i.e. $k$) in the oscillation quenching and transition scenarios. From Eq.(\ref{pbeps}) and Eq.\eqref{hopfeps} it is clear that ${\epsilon}_{HB}$ and ${\epsilon}_{PB}$ collide at $k=2$, thus the AD region (and hence AD-OD and AD-NAD transitions) vanishes for $k\le2$. Fig.\ref{fig4} shows this in the $\epsilon-k$ parameter space for a fixed $d$ (we choose $d=1.5$ such that $d_{HB}<d<d_{PB}$). We can see that for $k>2$, AD-NAD transition occurs with varying $\epsilon$; also the zone of AD region surrounded by HB2 and PB2 curves gets narrower for decreasing $k$, and vanishes at $k=2$, which is in accordance with our earlier result. Since $d_{HB}$ and $d_{PB}$ both are independent of $k$, thus variation of $k$ does not affect the dynamics that is entirely controlled by $d$.

\subsection{Van der Pol oscillators with direct-indirect coupling}
Next, to verify the generality of the above oscillation quenching transitions in periodic oscillator, we consider two identical Van der Pol oscillators under the same coupling scheme; mathematical model of the coupled system is given by
\begin{subequations}
\label{vdpsystem}
\begin{align}
\label{vx}
\dot{x}_{1,2} &= y_{1,2}+d(x_{2,1}-x_{1,2})+\epsilon s,\\
\label{vy}
\dot{y}_{1,2} &= a(1-{x_{1,2}}^2)y_{1,2}-x_{1,2},\\
\label{vs}
\dot{s} &= -ks-\frac{\epsilon (x_1+x_2)}{2}.
\end{align}
\end{subequations}
From Eq.(\ref{vdpsystem}) we can see that the origin $(0,0,0,0,0)$ is the homogeneous steady state (HSS), and also we have  two coupling dependent nontrivial fixed points, $\mathcal{F}_{IHSS}\equiv(x^\dagger, y^\dagger, -x^\dagger, -y^\dagger, 0)$, where 
$x^\dagger = \frac{y^\dagger}{2d}$, $y^\dagger = \pm \sqrt{4d^2-\frac{2d}{a}}$ and $\mathcal{F}_{NHSS}\equiv(x^\ast, y^\ast, x^\ast, y^\ast, s^\ast)$, where 
$x^\ast = \frac{ky^\ast}{{\epsilon}^2}$, $y^\ast = \pm \sqrt{\frac{{\epsilon}^2(a{\epsilon}^2-k)}{ak^2}}$ and $s^\ast = -\frac{\epsilon x^\ast}{k}$. 

\begin{figure}
\label{tpbdvdp}
\includegraphics[width=.48\textwidth]{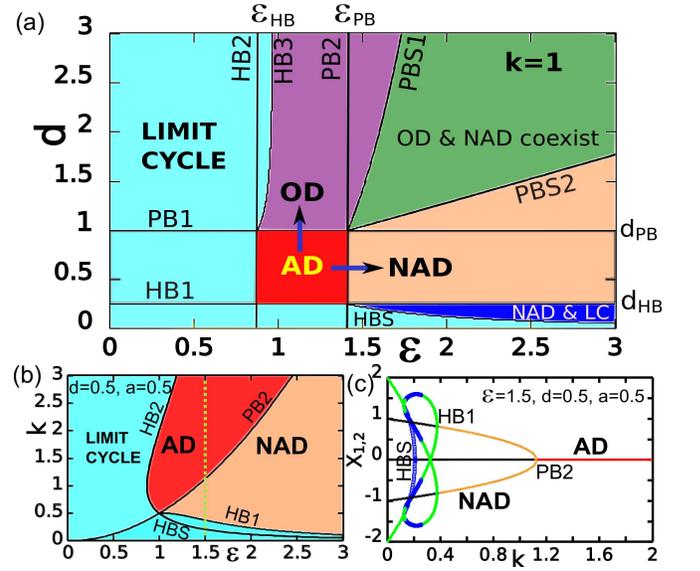}
\caption{\label{fig5} (Color online)(a) Two-parameter bifurcation diagram in $\epsilon - d$ space for Van der Pol oscillator ($k=1$). AD-NAD and AD-OD transitions are indicated with arrow. (b) Two-parameter bifurcation diagram in $\epsilon - k$ space ($d = 0.5$). (c) One dimensional bifurcation diagram with $k$ at $\epsilon=1.5$ shows a transition among the LC-NAD-AD. Other parameter: $a = 0.5$.}
\end{figure}

The characteristic equation of the system at the fixed point $(0,0,0,0,0)$ is given by,
\begin{equation}
\label{vtrieigen}
({\lambda}^2+{{Q'}}_{T1}\lambda + {{Q'}}_{T0})({\lambda}^3+{Q}_{T2}{\lambda}^2+{Q}_{T1}\lambda +{Q}_{T0})=0,
\end{equation}
where ${{Q'}}_{T1}=(2d-a)$, ${{Q'}}_{T0}=(1-2ad)$, ${Q}_{T2}=(k-a)$, ${Q}_{T1}=(1+{\epsilon}^2-ka)$, ${Q}_{T0}=(k-{\epsilon}^2a)$. Like the Stuart-Landau oscillator case, $\mathcal{F}_{IHSS}$ and $\mathcal{F}_{NHSS}$ emerge through pitchfork bifurcation at $d_{PB}=\frac{1}{2a}$ and $\epsilon_{PB} = \sqrt{\frac{k}{a}}$, respectively.
Also, using the similar arguments used in the previous subsection, we derive two Hopf bifurcation points from Eq.\eqref{vtrieigen} as
\begin{equation}
\label{dhbvdp}
d_{HB} = \frac{a}{2},
\end{equation}
\begin{equation}
\label{hb1vdp}
{\epsilon}_{HB} = \sqrt{\frac{a(1+k^2)-ka^2}{k}}.
\end{equation}
Two-parameter bifurcation diagram using $\mbox{XPPAUT}$ is shown in Fig.\ref{fig5}(a) (for $k=1$, $a = 0.5$). As ${\epsilon}_{PB}$ and ${\epsilon}_{HB}$ collide at $k=a$, thus AD region vanishes for $k\le a$.  We identify the following two distinct transition scenarios (for $k>a$): (i) If ${\epsilon}_{HB}<\epsilon<{\epsilon}_{PB}$, variation of $d$ gives rise to a transition from limit cycle (LC) to AD (at $d_{HB}$), and from AD to OD (at $d_{PB}$). (ii) If $d_{HB}<d<d_{PB}$, variation of $\epsilon$ induces a transition from LC to AD (at $\epsilon_{HB}$), and from AD to NAD (at $\epsilon_{PB}$). 

Outside this region, we have identified other bifurcation curves which are qualitatively equivalent to that of the  Stuart-Landau oscillator. The characteristic equations for the nontrivial fixed points is given by
\begin{equation}
\label{vntrieigen}
({\lambda}^2+{{Q'}^i}_{N1}\lambda + {{Q'}^i}_{N0})({\lambda}^3+{Q^i}_{N2}{\lambda}^2+{Q^i}_{N1}\lambda +{Q^i}_{N0})=0,
\end{equation}
where ${{Q'}^i}_{N1}=a({x^i}^2-1)+2d$, ${{Q'}^i}_{N0}=2ax^iy^i+1-2da(1-{x^i}^2)$, ${Q^i}_{N2}=k+a({x^i}^2-1)$, 
${Q^i}_{N1}=1+{\epsilon}^2+2ax^iy^i-ka(1-{x^i}^2)$, ${Q^i}_{N0}=k(1+2ax^iy^i)-{\epsilon}^2a(1-{x^i}^2)$. Where $i$ will be replaced by $\dagger$ and $\ast$ for the nontrivial fixed points $\mathcal{F}_{IHSS}\equiv(x^\dagger, y^\dagger, -x^\dagger, -y^\dagger, 0)$ and $\mathcal{F}_{NHSS}\equiv(x^\ast, y^\ast, x^\ast, y^\ast, s^\ast$), respectively. The locus of different bifurcation curves are derived from Eq.\eqref{vntrieigen} using the same method discussed in the coupled Stuart-Landau oscillators and they are given by
\begin{subequations}
\begin{align}
\epsilon_{HBS} &= \sqrt{\frac{k}{2d}},\\
d_{HB3}&= \frac{(1-k^2)+\sqrt{(k^2-1)^2+k(8a-4k{\epsilon}^2)}}{8a-4k{\epsilon}^2},\\
\epsilon_{PBS1} &= \sqrt{\frac{k+\sqrt{k^2(1+16ad)}}{4a}},\\
d_{PBS2} &= \frac{k+\sqrt{k^2+8ak{\epsilon}^2}}{8ak}.
\end{align}
\end{subequations}
All the oscillation quenching scenarios remain qualitatively same as that of the Stuart-Landau (SL) oscillator except now the HBS1 curve of SL case is replaced by HB3 curve. We also show the two-parameter bifurcation diagram in $\epsilon-k$ space [Fig.\ref{fig5}(b)], which depicts that for $k<a$ no AD is possible (here $a=0.5$ and $d=0.5$). The corresponding one dimensional bifurcation diagram with $k$ at $\epsilon=1.5$ is shown in Fig.\ref{fig5}(c) that shows a transition from limit cycle to NAD to AD state (later it is confirmed in the experiment, also).
\section{Experiment}
\label{sec:expt}
\begin{figure}
\includegraphics[width=.43\textwidth]{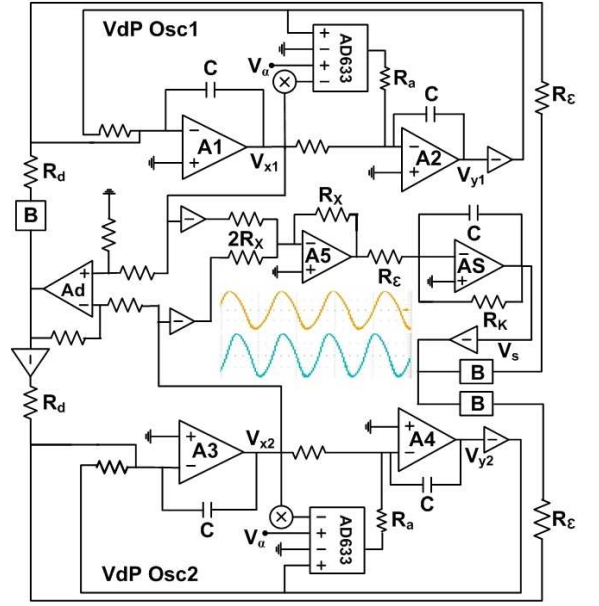}
\caption{\label{ckt} (Color online) Experimental circuit diagram of directly and indirectly coupled VdP oscillators. AS, Ad, A1-A5 op-amps are realized with TL074 (JFET). All the unlabeled resistors have value $R=10$~k$\Omega$. C=10 nF, $R_a=200\Omega$, $R_X=10$~k$\Omega$, $R_k=10$~k$\Omega$ ,$V_\alpha=0.1$ v. $\pm12$ v power supplies are used; resistors (capacitors) have $\pm5\%$ ($\pm1\%$) tolerance. Box denoted by ``B" are op-amp based buffers; inverters are realized with the unity gain inverting op-amps. $\otimes$ sign indicates squarer using AD633. Inset (in the middle part) shows the  oscillation from the uncoupled VdP oscillators: upper trace (yellow) $V_{x1}$, lower trace (cyan) $V_{x2}$ ($y$-axis:10 v/div, $x$-axis:500 $\mu$s/div).}
\end{figure}

We experimentally verify the occurrence of all the transitions predicted in the above section. For this purpose we implement an electronic circuit that mimics the coupled Van der Pol oscillators \cite{vdpckt} with the direct-indirect coupling given by Eq. \eqref{vdpsystem}; Fig.~\ref{ckt} depicts the said circuit. In this circuit diagram the sub unit associated with the op-amp ``Ad'' acts as an differential amplifier and emulates the diffusive coupling part. The sub circuit of the op-amp ``AS'' mimics the environmental part whose damping parameter ($k$) is controlled by the resistor $R_k$. The voltage equation of the circuit of Fig.~\ref{ckt} can be written as:
\begin{subequations}
\label{ckteq}
\begin{align}
\label{xckt}
CR\frac{dV_{xi}}{dt}&= V_{yi}+\frac{R}{R_d}\bigg(V_{xj}-V_{xi}\bigg)+\frac{R}{R_{\epsilon}}V_s,\\
\label{yckt}
CR\frac{dV_{yi}}{dt}&= \frac{R}{100R_a}\bigg(10V_{\alpha}-{V_{xi}}^2\bigg)V_{yi}- V_{xi},\\
\label{sckt}
CR\frac{dV_s}{dt}&= -\frac{R}{R_k}V_s-\frac{R}{R_{\epsilon}}\bigg(\frac{V_{xi}+V_{xj}}{2}\bigg).
\end{align}
\end{subequations}
Here, $i,j=1,2$ and $i\neq j$. Equation (\ref{ckteq}) is normalized with respect to $CR$, and thus now becomes equivalent to Eq. (\ref{vdpsystem}) for the following normalized parameters: $\dot{u}=\frac{du}{d\tau}$, $\tau = \frac{t}{CR}$, $\epsilon = \frac{R}{R_{\epsilon}}$, $d=\frac{R}{R_d}$, $k=\frac{R}{R_k}$, $a=\frac{R}{100R_a}$, $10V_{\alpha}=1$, $x_i=\frac{V_{xi}}{V_{sat}}$, $y_i=\frac{V_{yi}}{V_{sat}}$, and $s=\frac{V_s}{V_{sat}}$. Thus, $R_d$ determines the diffusive coupling strength $d$ and $R_{\epsilon}$ determines the environmental coupling strength $\epsilon$.  $V_{sat}$ is the op-amp saturation voltage. To make our circuit equivalent to Eq. \eqref{vdpsystem} we take $V_{\alpha}=0.1$; also, $a = 0.5$ (i.e. $R_a= 200$ $\Omega$) is taken to match with the parameter value used in Fig.\ref{fig5}; $V_{\alpha}$ and $a$ determine the amplitude and shape of the limit cycle. Also, we choose $C = 10$~nF and $R=10$~k$\Omega$ that determine the frequency of individual oscillations, which is, in this case, $1.54$ kHz (for the uncoupled case), and are shown in Fig.~\ref{ckt} (inset). We experimentally verify that the choice of $V_{\alpha}$, $a$ and $\tau$ (i.e., $C$ and $R$) does not affect the qualitative features of the {\it coupled dynamics}.

\begin{figure}
\includegraphics[width=.5\textwidth]{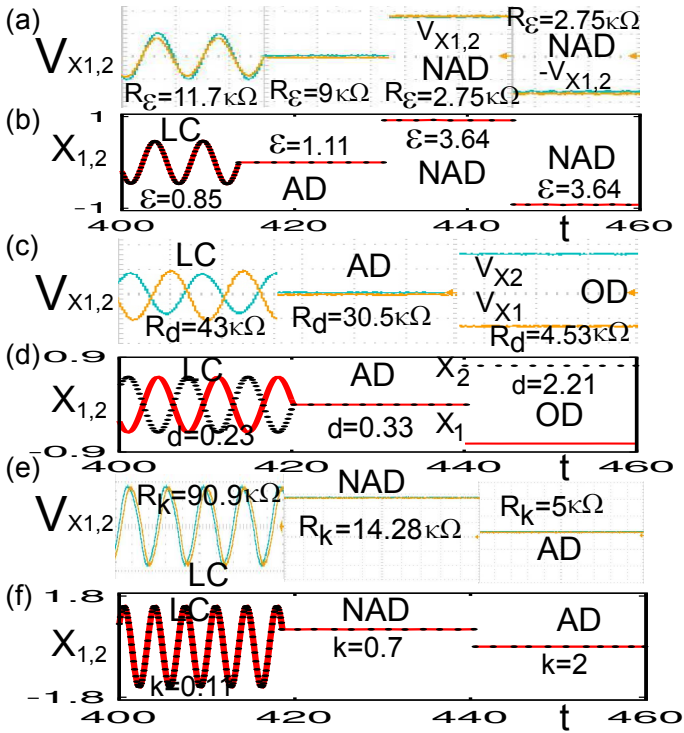}
\caption{\label{expt} (Color  online)  Experimental real time traces [(a),(c) and (e)] of $V_{x1}$ and $V_{x2}$ along with the numerical time series plots [(b),(d) and (f)] of $x_1$ and $x_2$.  {\bf (a, b) Variation of environmental coupling} ($R_d=30$~k$\Omega$, $d=0.33$): limit cycle at $R_\epsilon=11.7$~k$\Omega$ ($\epsilon=0.85$), AD at $R_\epsilon=9$~k$\Omega$ ($\epsilon=1.11$), and NAD at $R_\epsilon=2.75$~k$\Omega$ ($\epsilon=3.64$). Two bistable NAD states $V_{x1,2}$ and $-V_{x1,2}$ at $R_\epsilon=2.75$~k$\Omega$ ($\epsilon=3.64$) are shown. {\bf (c, d) Variation of diffusive coupling} ($R_{\epsilon}=8.33$~k$\Omega$, $\epsilon=1.2$): limit cycle at $R_d = 43$~k$\Omega$ ($d=0.23$), AD at $R_d = 30.50$ ~k$\Omega$ ($d=0.33$) and OD state $R_d=4.53$~k$\Omega$ ($d=2.21$).  {\bf (e, f) Variation of environment ($k$)} [$R_d=20$k$\Omega$ ($d=0.5$), $R_{\epsilon}=6.66$k$\Omega$ ($\epsilon=1.5$)]: limit cycle at $R_k = 90.9$~k$\Omega$ ($k=0.11$), NAD at $R_k = 14.28$ ~k$\Omega$ ($k=0.7$) and AD at $R_k=5$~k$\Omega$ ($k=2$); for clarity only the $V_{x1,2}$ NAD state is shown. [(a),(c) and (e): $y$-axis: 5 v/div, $x$-axis: 500 $\mu$s/div, for (e) $x$-axis: 250 $\mu$s/div].}
\end{figure}

In this experiment at first we take $R_d=30$~k$\Omega$ ($d=R/R_d=0.33$, i.e., $d_{HB}<d<d_{PB}$) and $k=1$ (i.e. $R_k = 10$ k$\Omega$): we observe a continuous transition from limit cycle to AD, and {\it AD to NAD} for decreasing $R_{\epsilon}$ (i.e. increasing $\epsilon$). We notice that in the limit cycle state two systems are in complete synchronized state. In Fig.~\ref{expt}~(a), using the experimental snapshots of the waveforms [with a digital storage oscilloscope (Tektronix TDS2002B, 60 MHz, 1 GS/s)], we demonstrate different dynamical behaviors for the following parameter values: limit cycle at $R_\epsilon=11.7$~k$\Omega$, AD for $R_\epsilon=9$~k$\Omega$, and NAD for $R_\epsilon=2.75$~k$\Omega$. Two bistable NAD states $V_{x1,2}$ and $-V_{x1,2}$ at $R_\epsilon=2.75$~k$\Omega$ are found by random parameter sweeping around $R_\epsilon=2.75$~k$\Omega$. For the comparison purpose, we also show the numerical results (using the fourth-order Runge-Kutta method, $0.01$ step size) taking $\epsilon$ and $d$ values that are equivalent to $R_\epsilon$ and $R_d$, respectively: for $d = 0.33$, Fig.\ref{expt}~(b) shows the limit cycle for $\epsilon=0.85$, AD for $\epsilon=1.11$, and {\it bistable} NAD for $\epsilon=3.64$ (using proper initial conditions). Next, we take $R_{\epsilon}=8.33$~k$\Omega$ (i.e., $\epsilon=R/R_\epsilon=1.2$; ${\epsilon}_{HB}<\epsilon<{\epsilon}_{HB}$) and observe a continuous transition from limit cycle to AD, and AD to OD for decreasing $R_d$. Fig.~\ref{expt}~(c), and Fig.~\ref{expt}~(d), respectively demonstrate experimental and numerical results: limit cycle (anti-phase synchronized) for $R_d = 43$~k$\Omega$ ($d=0.23$), AD for $R_d = 30.50$ ~k$\Omega$ ($d=0.33$), and OD for $R_d=4.53$~k$\Omega$ ($d=2.21$). 

Next, we experimentally verify the effect of the environment (i.e. $k$) on the coupled dynamics. For this we choose the following circuit parameters: $R_d=20$k$\Omega$ (i.e., $d=0.5$) and $R_{\epsilon}=6.66$k$\Omega$ (i.e., $\epsilon=1.5$) and vary $R_k$. With decreasing $R_k$ (i.e., increasing $k$) we observe a transition from LC to NAD to AD, which is in accordance with the numerical result of Fig.\ref{fig5} (c). Figure \ref{expt} (e) and (f) show this scenario experimentally and numerically, respectively (see caption for details). Thus, we see that in a real system, despite the presence of inherent noise, parameter fluctuation and mismatch, the experimental observations are qualitatively similar to our theoretical results.

\section{Conclusion}
\label{sec:con}
We conclude that although the simultaneous application of direct and indirect coupling was proposed as a general scheme for inducing amplitude death state \cite{resmiad}, in this paper we have shown that this coupling scheme can induce several other oscillation quenching states such as OD and a new nontrivial AD (NAD) state that is a bistable state; this NAD state has different manifestation and genesis compared to the conventional AD state. Further, we have shown that this coupling scheme can induce transitions from AD to OD (with the  variation in diffusive coupling strength), and AD to NAD (with the variation in environmental coupling strength). The latter transition is reported for the first time. We have also reported the first experimental evidence of oscillation suppression states and their transitions induced by this direct-indirect coupling. 

In the present paper we have investigated two generic periodic oscillators that have certain type of symmetry. In biological systems (where negative concentrations are forbidden) and chaotic systems the bifurcation scenarios will be much more complex. Since in biological oscillators non-negative asymmetric OD states have already been reported (see e.g. Ref.\cite{kosprep}), we hope that the theoretical findings of this paper will be valid for biological oscillators, also. But, the exact bifurcation scenario may be different and that requires further investigations. Further, we test the system dynamics for a regular network of large number of oscillators (not reported in the present manuscript) and find that the said AD, OD and NAD states are preserved for the larger system, also.

Since in biological systems both diffusive and environmental coupling are omnipresent \cite{resmi12}, thus we believe that this detailed study will improve our understanding of several biological processes. More specifically, as we have shown that in the NAD state the system becomes bistable, thus, further research is required to explore any possible connection of NAD to the bistability arises in many biological processes such as the bistability in the brain activity \cite{brain,*brain2} (e.g. sleep-wake cycle of mammals and birds), {\it lac} operon in the bacteria {\it E. coli} \cite{ecoli}, cell cycle \cite{cell2}, etc. Further, one of the central topics of recent research is to identify the coupling schemes that can implement bistability in biological systems and many schemes have already been identified  and experimentally verified in this context \cite{new}. Future research can be undertaken in order to explore the role of the direct-indirect coupling studied here in inducing bistability in biological systems.

\acknowledgments{Authors are thankful to the anonymous referees for their constructive comments and suggestions. T.B. acknowledges the financial support from SERB, Department of Science and Technology (DST), India [project grant: SB/FTP/PS-005/2013]. D.G. acknowledges DST, India for providing the INSPIRE fellowship.}

\providecommand{\noopsort}[1]{}\providecommand{\singleletter}[1]{#1}%
\end{document}